\documentstyle[12pt]{article}
\newcommand{\rf}[1]{(\ref{#1})}
\newcommand{\bea}{\begin{eqnarray}}
\newcommand{\eea}{\end{eqnarray}}

\renewcommand{\l}{\lambda}
\renewcommand{\b}{\beta}

\newcommand{\n}{\nu}
\newcommand{\m}{\mu}

\newcommand{\sg}{\sigma}

\newcommand{\oh}{\frac{1}{2}}

\newcommand{\dg}{\dagger}

\newcommand{\tr}{{\rm Tr}\;}
\newcommand{\ra}{\right\rangle}
\newcommand{\la}{\left\langle}
\newcommand{\sgn}{{\rm sgn}}

\newcommand{\Vqq}{V_{q\bar{q}}}
\def\void{}
\def\labelmark{}

\newenvironment{formula}[1]{\def\labelname{#1}
\ifx\void\labelname\def\junk{\begin{displaymath}}
\else\def\junk{\begin{equation}\label{\labelname}}\fi\junk}%
{\ifx\void\labelname\def\junk{\end{displaymath}}
\else\def\junk{\end{equation}}\fi\junk\labelmark\def\labelname{}}

{\ifx\void\labelname\def\junk{\end{array}\end{displaymath}}
\else\def\junk{\end{array}\right.\end{equation}}
\fi\junk\labelmark\def\labelname{}\def\junk{}
}

\newcommand{\beq}{\begin{formula}}
\newcommand{\eeq}{\end{formula}}
\newcommand{\beqv}{\begin{formula}{}}

\begin{document}
\topmargin 0pt
\oddsidemargin 5mm
\headheight 0pt
\headsep 0pt
\topskip 9mm

\hfill    NBI-HE-94-30

\hfill March 1994

\begin{center}
\vspace{24pt}
{\large \bf Scaling with a modified Wilson action which suppresses\\
 Z$_2$ artifacts in SU(2) lattice gauge theories}

\vspace{24pt}

{\sl J. Ambj\o rn } and  {\sl G. Thorleifsson}

\vspace{6pt}

 The Niels Bohr Institute\\
Blegdamsvej 17, DK-2100 Copenhagen \O , Denmark\\

\end{center}

\vspace{24pt}

\addtolength{\baselineskip}{0.20\baselineskip}
\vfill

\begin{center}
{\bf Abstract}
\end{center}

\vspace{12pt}

\noindent
A modified Wilson action which suppresses plaquettes which take
negative values is used to study the scaling behavior of the
string tension. The use of the $\b_E$ scheme gives good agreement
with asymptotic two loop results.

\vfill

\newpage

\section{Introduction}
The simplest and most popular choice of a gauge lattice action is
the one proposed by Wilson, which for the group $SU(2)$ reads:
\beq{*1}
S_W (U_l) = \b \sum_{\Box} (1 -\oh \tr U_\Box ),
\eeq
where $\b = 4/g^2$, $U_l \in SU(2)$ is the link variable
defined on the link $l \equiv (x,\m)$, while $\Box \equiv (x,\m\n)$
refers to the location and orientation of the corresponding plaquette.
$U_\Box$ is the standard plaquette variable:
\beq{*2}
U_\Box \equiv U_{x,\m\n} = U_{x,\m}U_{x+\m,\n}U^\dg_{x+\n,\m}U^\dg_{x,\n}.
\eeq

In any Monte Carlo simulation the crucial question is whether the
field configurations generated provide an adequate representation
of the continuum physics. The naive continuum limit can only
be obtained for $\oh \tr U_\Box \approx 1$. Especially
any lattice configurations which locally have some $U_\Box$'s where
$\oh \tr U_\Box \approx -1$  are lattice artifacts which a priori
have nothing to do with continuum configurations. A complete suppression
of such configurations should not change any continuum physics. For
the values of $\b$ where one performs the Monte
Carlo simulations such local $Z_2$ fluctuations are not at all rare if
$S_W$ is used. It is thus very important to make sure that these fluctuations
have no impact on continuum observables. The easiest way to do so is to
modify the lattice action in such a way that these small-scale fluctuations
are suppressed, without influencing plaquettes where $\tr U_\Box > 0$.
This can be done by modifying the Wilson action as follows \cite{bcm}:
\beq{*3}
S=S_W+S_\l,
\eeq
where $S_W$ is the standard Wilson action \rf{*1}, while $S_\l$ is
\beq{*4}
S_\l = \l \sum_{\Box} [1-\sgn( \tr U_\Box)].
\eeq

The action \rf{*3} was studied for $\l =\infty$ in \cite{mp} while
the phase structure in the $(\b,\l)$ plane was studied in \cite{bcm}.
For a fixed $\l$ one would expect the same continuum limit for
$\b \to \infty$ and also the same identification $\b = 4/g^2$.
However, as shown in \cite{bcm} there is a marked difference
between, say, Creutz ratios, measured for $\l =0$ (the Wilson action)
and for $\l =0.5$ in the case of $\b=2.5$.
One would expect the difference to be even more pronounced with increasing
$\l$. This illustrates that plaquettes with negative values may
play an important role for the range of $\b$'s where the scaling of
pure $SU(2)$ lattice gauge theories is usually studied, and one could
be worried about the relation to continuum physics. The fact that one gets
acceptable agreement with continuum scaling relations could be fortuitous
since one would expect the action $S$ to reproduce continuum physics
better for large $\l$. One  purpose of this article is to show that
we indeed get the correct scaling relations for large $\l$ and that there
is no reason to expect the plaquettes with negative values which appear
in the Wilson action to play any important role for $\b \geq 2.2$.
Another purpose is to test whether the $\l$ modification of the action
may improve the approach to scaling for the reasons mentioned above.

\section{Numerical method}

The numerical simulations were performed using a standard
Metropolis algorithm to update the gauge fields, combined
with an overrelaxation algorithm  to
decrease the autocorrelations.  For every Metropolis sweep
we performed 4 overrelaxation steps.
Lattice size
$12^4$ with periodic boundary conditions was used and we
measured
all Wilson loops up to the size $6 \times 6$.
In order to improve the statistics we modified the
configurations using the Parisi trick
\cite{ppf} before measuring.  Unlike in the case of
$\lambda = 0$ the integration involved in the Parisi trick
had to be performed numerically and was thus costly in
computer time.  But the gain in the statistics did more than
compensate for that.

The choice of parameters was $\lambda = 10$, which was
sufficient to suppress all negative plaquettes, and
$\beta \in [1.0,2.5]$.
We found a scaling window for $\beta \in
[1.5,1.9]$ and concentrated the simulations in
that interval.  Usually 1000 sweeps where used for
thermalization and the number of sweeps used
for measuring, for various $\beta$,
is shown in table 1.  Measurements were performed
every fifth sweep and the errors were estimated by
using the jackknife method.

\begin{table}
\begin{center}
\begin{tabular}{c|c|c}
$\beta$   & No. of sweeps   & No. of measurements   \\  \hline
1.5    &   55250    &   11050  \\
1.6    &   65250    &   13050  \\
1.7    &   63750    &   12750  \\
1.8    &   88750    &   17750  \\
1.9    &   45000    &   9000

\end{tabular}
\caption{Statistics collected at various $\beta$. In all measurements
$\lambda = 10$}
\end{center}
\end{table}

\section{Scaling of the string tension}
The $\b$-values in the scaling window are even smaller
than the ones used for the standard Wilson action. If one wants to
test scaling by comparing  with the perturbative two-loop result
using the identification $\b = 4/g^2$
it is doomed to fail. We clearly need a suitable  effective coupling
which we can use instead of $g^2$. Here we will use the
so-called $\b_E$ scheme  \cite{parisi,kp,smm}. This scheme is simple
and  well suited to deal with ``perturbations'' of the Wilson
action, like the ones given by \rf{*3}-\rf{*4}. Let us define
\beq{*5}
\b_E = \frac{3}{4} \frac{1}{1-\la \oh \tr U_\Box \ra}.
\eeq
{}From weak coupling expansions ($\b \to \infty,~~\l$ fixed), it is seen that
\beq{*6}
\b_E \to \b~~~~~{\rm for}~~~~ \b \to \infty.
\eeq
It is consistent to use the $g_E=4/\b_E$ in the two loop $\b$-function
since $g_E$ is a function of $g$ and the two first coefficients of the
$\b$-function are invariant under a non-singular coupling constant
redefinition. The use of $\b_E$ in the two loop $\b$-function
is known to diminish scaling violations in a variety of situations:
The string tension in $SU(2)$ and $SU(3)$ gauge theories,
the mass gap in the  $SU(2)$ and $SU(3)$  chiral models and the
deconfining transition temperature $T_c$, again for $SU(2)$ and $SU(3)$
gauge theories, just to mention some. We will now show that the method
works well for the modified Wilson action with $\l=10$.

Let us first note that for $\l =10$ the $\b$-range $1.4-1.9$ is mapped
into the $\b_E$-range $1.837-2.104$, which should compared to the
corresponding map for the ordinary Wilson action where the
$\b$-range $2.2-2.5$ is mapped into the $\b_E$-range $1.741-2.154$.
{}From the measurements of Wilson loops mentioned in the last section
we extract the string tension using the simplest method: If $W(R,T)$
denotes a $R \times T$ Wilson loop we first form
\beq{*7}
V_{eff} (R,T) = \log [W(R,T)/W(R,T-1)].
\eeq
For $T \to \infty$ $V_{eff} (R,T)$ should go to the ground state static
quark potential $\Vqq(R)$. We have available $T \leq 6$ and the results
agree within error bars for $T=5-6$. We have used these values as $\Vqq(R)$
for $R \leq 6$. Next $\Vqq(R)$ is fitted to the form:
\beq{*7a}
\Vqq(R) = C - \frac{E}{R} + \sg R
\eeq
and  the error bars for $\sg$ are mainly coming from
systematic errors arising
from fits with various lower cuts in $R$. It is clear that with the
limited range of $R$ and $T$'s available to us nothing will be gained
by applying some of the many more elaborate schemes of fitting which are
available.

$\log \sg(\b_E)$ is finally plotted in fig. 1 against $\b_E$. The curve
shown is the one obtained from the two-loop $\b$-function, according to
which the scaling of the lattice string tension
$\sg(\b_E) = \sg_{cont} a^2(\b_E)$ is governed by the two-loop scaling
of the lattice spacing $a(\b_E)^2$:
\beq{*8}
 a^2(\b_E) = \Lambda^{-2}
\left( \frac{6\pi^2}{11} \b_E\right)^{102/121}
\exp\left[-\frac{6\pi^2}{11} \b_E\right].
\eeq
It is clear that scaling is reasonable well satisfied in the given
range of $\b_E$.

It is interesting to compare these results with the corresponding
ones obtained by using the ordinary Wilson action. The range of $\b$
which gives the same range of $\b_E$ as were used for the
modified action will be $2.3 \le \b \le 2.5$. In this range it is
well known that naive application of scaling does not work particularly
well, and to get a fair comparison with the results for the modified
action we use again the $\b_E$ coupling constant transformation.
We now follow the same method as in the case of the modified action
and extract the string tension\footnote{The data used
for $\b=2.4$ and 2.5 are taken from \cite{gutbrod}.}. The result is
shown as a function of $\b_E$ in fig. 1. The use of $\b_E$ has
improved the scaling considerable compared to
the use of $\b$ as a coupling constant, as already mentioned, and it
is apparent from fig.1 that agreement with the two-loop scaling
is as good for the Wilson action  as for the modified action.

\section{Conclusions}
The suppression of negative valued plaquettes should not change the
continuum limit of pure $SU(2)$ lattice gauge theories since these
configurations are pure lattice artifacts. Naively one might even
expect a smoother approach to the continuum limit if such a suppression
is implemented.  The modified Wilson action \rf{*3} indeed suppresses
negative valued plaquettes for large values of $\l$ and it is possible
to approach continuum physics for much smaller values of $\b$. However,
for these small values of $\b$ the relation between the continuum
coupling constant $g^2$ and $\b$ is not a simple one and in order to
extract the scaling behavior one has to use a modified coupling constant
closer reflecting the physics of the system. The $\b_E$-scheme is such
a prescription and using it we have found good agreement with continuum
physics. From these results it seems clear that
the theory with modified Wilson action
belongs to the same universality class as the theory defined by the
ordinary Wilson action when $\b$ is sufficiently large. The worry
mentioned in \cite{bcm} is thus ruled out. In addition it seems
that the approach to scaling is not dramatically improved
compared to the situation for ordinary Wilson action. Indirectly
this indicates that negative plaquette excitations play no important
role in the scaling region for the ordinary Wilson action, at least
when we discuss observables like the string tension. However,
one would expect that the modified Wilson action is
considerable better when it comes to the measurements of
topological objects like instantons.

\vspace{24pt}

\noindent
{\bf Acknowledgement} We  thank V. Mitrjushkin for discussions and
many useful suggestions. Part of the computations were performed
at UNI-C and were made possible by a grant from the
Danish Ministry of Research and Technology.

\vspace{24pt}

\addtolength{\baselineskip}{-0.20\baselineskip}

\vspace{36pt}

\noindent
{\bf Figure 1:} The measured string tension  using the modified
action (upper curve) and the ordinary Wilson action (lower curve)
as functions  of $\beta_E$.

\end{document}